\documentclass{pasa}%
\pdfoutput=1
\usepackage[usenames, dvipsnames]{color}
\title[The XXL-South redshift catalogue]{The XXL survey \\ XIV. AAOmega redshifts for the southern XXL field}
\author[C. Lidman et al.]{C. Lidman$^1$, F. Ardila$^{1,2}$,
  M. Owers$^{1,3}$, C. Adami$^4$,  L. Chiappetti$^5$, F. Civano$^{6,13}$,
  A. Elyiv$^{7,8}$, F. Finet$^{9,10}$,
  S. Fotopoulou$^{11}$,A. Goulding$^{11,12}$, E. Koulouridis$^{14}$,
  O. Melnyk$^{15,20}$, F. Menanteau$^{16,17}$, F. Pacaud$^{18}$,
  M. Pierre$^{14}$, M. Plionis$^{19}$,  J. Surdej$^{9}$,
  T. Sadibekova$^{14}$\\
\\
\affil{$^1$ Australian Astronomical Observatory, North Ryde, NSW 2113, Australia. e-mail clidman@aao.go.au}
\affil{$^2$ Department of Astronomy, University of Florida, Gainesville, FL 32611, USA}
\affil{$^3$ Department of Physics and Astronomy, Macquarie University, NSW 2109, Australia}
\affil{$^4$ LAM, OAMP, Universit\'{e} Aix-Marseille CNRS, Pole de l'Etoile, Site de Ch\^{a}teau Gombert, 38 rue Fr\'{e}d\'{e}ric Joliot-Curie, 13388, Marseille 13 Cedex, France}
\affil{$^5$ INAF, IASF Milano, via Bassini 15, I-20133 Milano, Italy}
\affil{$^6$ Yale Center for Astronomy and Astrophysics, 260 Whitney Avenue, New Haven, CT 06520, USA}
\affil{$^7$ Dipartimento di Fisica e Astronomia, Universit\`{a} di Bologna, viale Berti Pichat 6/2, I-40127 Bologna, Italy} 
\affil{$^8$ Main Astronomical Observatory, Academy of Sciences of Ukraine, 27 Akademika Zabolotnoho St., 03680 Kyiv, Ukraine}
\affil{$^9$ Extragalactic Astrophysics and Space Observations (AEOS), University of Li\`{e}ge, All\'{e}e du 6 Ao\^{u}t, 17 (Sart Tilman, Bât. B5c), 4000 Li\`{e}ge, Belgium}
\affil{$^{10}$Aryabhatta Research Institute of Observational Sciences (ARIES), Manora Peak, Nainital-263 129, Uttarakhand, India}
\affil{$^{11}$ Department of Astronomy, University of Geneva, ch. d'Ecogia 16, 1290 Versoix, Switzerland}
\affil{$^{12}$ Department of Astrophysical Sciences, Princeton University, Princeton, NJ 08544, USA}
\affil{$^{13}$ Harvard-Smithsonian Center for Astrophysics, 60 Garden Street, Cambridge, MA 02138, USA}
\affil{$^{14}$ Service d’Astrophysique AIM, CEA Saclay, F-91191 Gif sur Yvette}
\affil{$^{15}$ Astronomical Observatory, Taras Shevchenko National University of Kyiv,  Observatorna str. 3, 04053 Kyiv, Ukraine}
\affil{$^{16}$ National Center for Supercomputing Applications, University of Illinois at Urbana-Champaign, 1205 W. Clark St., Urbana, IL  61801, USA}
\affil{$^{17}$ Department of Astronomy, University of Illinois at Urbana-Champaign, W. Green Street, Urbana, IL 61801, USA}
\affil{$^{18}$ Argelander Institut fuer Astronomie, Universit\"{a}t Bonn,  D-53121 Bonn, Germany}
\affil{$^{19}$ Aristotle University of Thessaloniki, Department of Physics, Greece}
\affil{$^{20}$ Department of Physics, University of Zagreb, Bijenicka cesta 32, HR-10000 Zagreb, Croatia}
}

\jid{PASA}
\doi{10.1017/pas.\the\year.xxx}
\jyear{\the\year}
\def\arcsec{\hbox{$^{\prime\prime}$}}

\begin{document}%
\begin{abstract} We present a catalogue containing the redshifts of
  3,660 X-ray selected targets in the XXL southern field. The
  redshifts were obtained with the AAOmega spectrograph and 2dF fibre
  positioner on the Anglo-Australian Telescope. The catalogue contains 1,515 broad line AGN,
  528 stars, and redshifts for 41 out of the 49 brightest X-ray
  selected clusters in the XXL southern field. 

%The targets consist of three main types:
%active galactic nuclei, galaxies in clusters and the field, and
%stars.

\end{abstract}
\maketitle%
\section{INTRODUCTION }
\label{sec:intro}

The past decade has seen a dramatic growth in the number of surveys
covering large areas of the sky at multiple wavelengths. While these
surveys are primarily designed to answer a few key scientific
questions, often their greatest value are the data products they
provide for use by the broader astronomical community. The XXL
survey\footnote{http://irfu.cea.fr/xxl} is one such survey.

The XXL survey \cite{Pierre2015} has used the XMM-Newton X-ray
telescope to image 50\,deg$^2$ with a sensitivity of
$5\times10^{-15}$\,ergs\,s$^{-1}$\,cm$^{-2}$ in the [0.5-2.0]\,kev
band.  XXL covers two similarly sized fields that are referred to as
the northern XXL field, which is centered on the XMM-LSS survey
\cite{Pierre2004}, and the southern XXL field, which is centered in a region
covered by the Blanco Cosmology Survey \cite{Desai2012}. The XXL
survey has detected hundreds of galaxy clusters out to $z\sim2$
\cite[hereafter, XXL Paper II]{Pacaud2015} and thousands of AGN out to $z\sim5$
\cite{Fotopoulou2015}.

The key scientific motivation behind the XXL survey,
  which is described in greater detail in Pierre et
  al. \shortcite{Pierre2015}, is to constrain cosmological parameters
  using the number density and clustering of X-ray selected
  clusters. Critical to the success of the survey are a
  well-characterised proxy for cluster masses \cite{Lieu2015}, and
  redshifts for the clusters.

The northern field has considerably better
spectroscopic coverage than the southern
field. Within the footprint of the northern field, many tens of
thousands of redshifts are publicly available from surveys such as
GAMA\footnote{http://www.gama-survey.org} \cite{Liske2015} and
VIPERS\footnote{http://vipers.inaf.it} \cite{Guzzo2014}, among
others. In comparison, there are many fewer
redshifts in the southern XXL field. The XXL collaboration has
therefore embarked on a program of follow-up spectroscopy and imaging
on multiple facilities in the Southern Hemisphere, including
spectroscopy with the Anglo-Australian Telescope (AAT), imaging with
DECam on the CTIO 4m Blanco Telescope (Gardner et al. in preparation),
and radio observations with the Australia Telescope Compact Array
\cite{Smolcic2015}.

In this paper, we publish the redshifts that were obtained during
three observing runs with the AAT. Around 3,600 redshifts were
obtained. The objects targeted broadly consist of three types: Active
Galactic Nuclei (AGN), which make up most of the objects, galaxies,
and stars.

The paper is structured as follows. In section 2, we
 define how targets were selected, summarise the
observations, and provide a detailed
description of how the data were processed. In the section 3, we
describe the redshift catalogue, and present a few graphs illustrating
the key statistics of the survey. Before concluding, we compute the
velocity dispersion for those clusters that have at least 10
spectroscopically confirmed members.  We then end with a summary of
the key aspects of the paper. Throughout our analysis we adopt the
AB magnitude system and the WMAP9 cosmology \cite{Hinshaw2013}
of $H_0 = 70$\,km/s/Mpc, $\Omega_M = 0.28$ and $\Omega_{\Lambda} = 0.72$.
% Need to check which system we are on

\section{Target Selection, Observations, and Data Processing}

Targets were selected according to criteria that were
  designed to select two kinds of objects. The first set of criteria were
designed to select AGN, while the second set were designed to select
galaxies in clusters and groups.

\subsection{AGN}

The AGN sample consists of about 4,500 X-ray point-like sources that
were within 6 arc-seconds of an optical source (16 $< r <$ 22 mag)
from the Blanco Cosmology Survey (BCS) \cite{Desai2012}.  We only
considered one counterpart for each X-ray source following the
procedure described in Chiappetti et al. \shortcite{Chiappetti2013}
and Melnyk et al. \shortcite{Melnyk2013}. The best ranked counterpart
had the highest probability $P_{XO}=\exp(-n(<r)d^{2})$ of being
associated with the X-ray source \cite{Downes1986}, where $d$
represents the angular distance between the X-ray source and its
counterpart, and $n(< r)$ is the density of objects brighter than the
r-band magnitude of the considered optical counterpart.
The median value of $P_{XO}$ for our sample is
  0.9969, with quartiles of 0.9907 and 0.9993. According to Tajer et
  al. \shortcite{Tajer2007} and Garcet et al. \shortcite{Garcet2007},
  counterparts with $P_{XO}>0.99$ are considered good and counterparts
  with $0.99 > P_{XO}>0.97$ are considered fair. We did not
  apply a lower bound to $P_{XO}$, but we did visually inspect all
  counterparts using the $g$, $r$ and $i$ images from the BCS. We
deleted from the list all bright stars, artefacts, and objects that
were fainter than our $r=22$ magnitude cut, which is roughly the limit
one can obtain redshifts in a few hours of exposure time on the AAT.

Compared to cluster galaxies, AGN were assigned a lower priority for
observation. Hence, there is a bias against observing AGN near
clusters. This bias needs to be taken into account when computing the
clustering of AGN. %This meant that not all 4,500 sources could be observed. 
%The number observed was 3,184. 
%Due to fibre collisions with cluster galaxies, which had higher
%priority (see below), and other objects in the AGN sample, 

\subsection{Cluster galaxies}

For the cluster sample, we first selected all objects brighter than
$r=22$ within a 500\,kpc radius of the cluster
centers. The angles on the sky were derived from the
  photometric redshifts of the
  clusters.\footnote{The photometric redshifts of
    the clusters span a large range, from $z=0.1$ to $z=1$. Over this
    range of redshifts, angles vary from 60 arc-seconds at high
    redshift to 270 arc-seconds at low redshift.} We then selected
objects that satisfy one of two conditions: i) they have a photometric
redshift that differs from the cluster redshift by 0.05 or less (0.05
is the median uncertainty in our photometric redshifts), or ii) they
have a colour that places them within the cluster red sequence. We
also manually inserted targets that were of special interest, e.g.~a
bright galaxy that was not inside the selected ranges listed above,
but was likely to be a cluster member.

\smallskip
\subsection{Observations}

We used the two-degree field (2dF) fibre positioner \cite{Lewis2002} on the AAT in
conjunction with the AAOmega spectrograph \cite{Smith2004} to observe 4,884 targets in
the XXL-South footprint during three observing runs in 2013. The dates of
the observing runs are listed in Table~\ref{tab:observingDates}.

\begin{table}[h]
\caption{Observing dates}
\begin{center}
\begin{tabular}{ll}
\hline\hline
July 05 to July 10 & six half nights \\
September 08 to September 09 & two full nights\\
September 29 & one full night \\
\hline\hline
\end{tabular}
\end{center}
\label{tab:observingDates}
\end{table}

The 2dF fibre positioner can place up to 400 fibres within a 2.1\,deg
diameter field-of-view, with the restriction that any two fibres
cannot be placed within 30\arcsec\ to 40\arcsec\ of each other. Eight fibres
are used for guiding, and following the recommendation in the AAOmega
User's
Manual,\footnote{http://www.aao.gov.au/science/instruments/AAOmega} 25
fibres were placed in areas free of objects as a means of obtaining a
spectrum of the night sky. This leaves of the order of 360 fibres that
can then be positioned on objects of interest. The fibre diameter on
sky is approximately 2\arcsec.

The 2dF fibre positioner can feed either HERMES \cite{Sheinis2015}, a moderate
resolution, 4-channel spectrograph, or the AAOmega spectrograph \cite{Smith2004}, which was
the instrument used in this study. A dichroic in AAOmega is used to
split the light into the red and blue channels of the spectrograph.  We
used one of the standard setups for our observations: the X5700
dichroic, the 385R grating in the red channel and the 580V grating in the
blue channel. This results in spectra that start and end at 370\,nm and
890\,nm, respectively, with a spectral resolution of about 1500. The
amount of overlap between the two channels varies with fibre number, and is typically 3\,nm

%Table~\ref{fieldCentres}

\begin{figure*}
\begin{center}
\includegraphics[width=40pc]{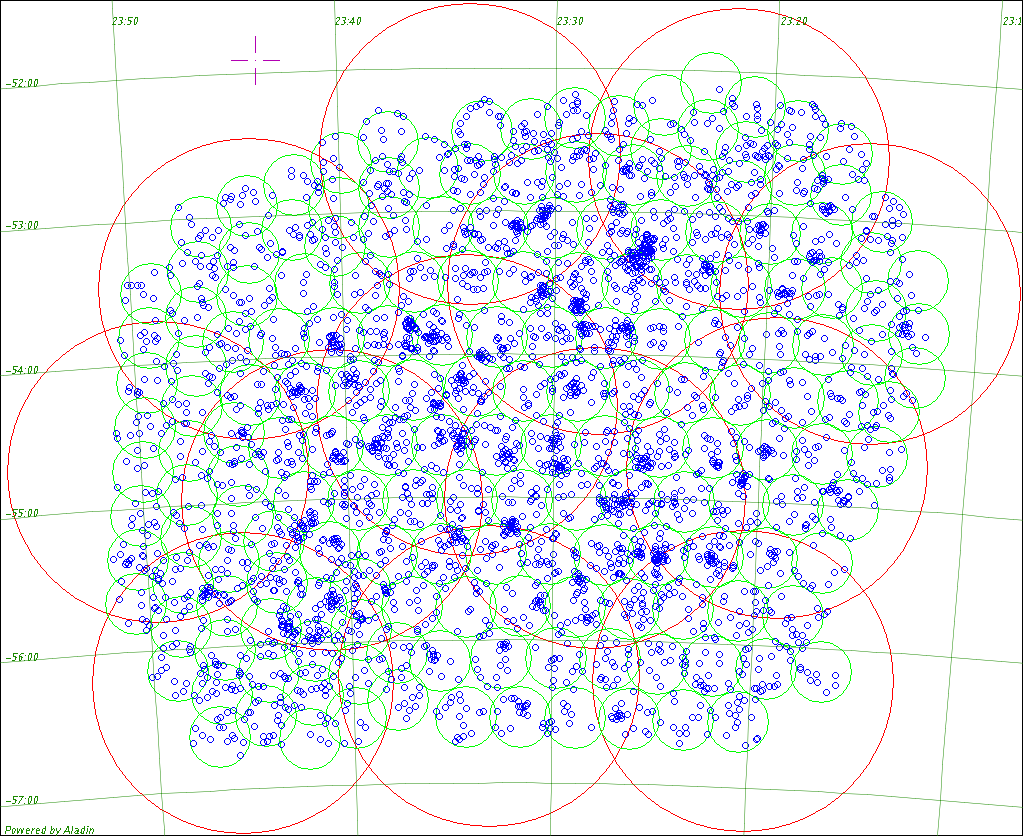}
\caption{A view of the objects (small blue circles) in the XXL
  southern field with spectroscopic redshifts from
    this work. The individual XMM pointings are shown as
    the green circles. The 13 larger red circles, each 2.1 degrees in
    diameter (the field of view of 2dF), cover the entire area
    observed with XMM.  Individual clusters can be picked out from
  the concentration of blue circles in certain areas. This plot was
  produced with Aladin.}
 \label{fig:objects}
\end{center}
\end{figure*}

The entire footprint of XXL-South is covered by 13 2dF fields. The
field centres of these 13 fields are %listed in Table~\ref{tab:fieldCentres} and 
shown in Fig.~\ref{fig:objects}. On average, three hours was spent on each
field over the course of the three observing runs.

When configuring each field, highest priority was given to candidate
cluster galaxies. Next highest priority were AGN, followed by sky
fibres. Science exposures lasted 30 minutes, and depending on observing
conditions, up to four of these were taken
consecutively in an observing sequence. Prior to each sequence, an arc
frame and a fibre flat were taken. The arc frame is used to calibrate
the wavelength scale of the spectra and the fibre flats are used to
determine the locations of the spectra on the CCD (the so-called
tramline map), to remove the relative wavelength dependent
transmission of the fibres (absolute normalisation is done using night
sky lines), and to determine the fibre profile for
later use in extracting the spectra from the science frames.

Bias frames and dark frames were taken daily. While these frames are
not required for processing data from the CCD in red arm of AAOmega,
they are essential for removing artefacts from data taken with the
CCD in the blue arm.

%\begin{table}
%\caption{AAT Field Centres}
%\begin{center}
%\begin{tabular}{cll}
%\hline\hline
%Field name & RA & Declination \\
 %          & (J2000) & (J2000) \\
%\hline%
% 1  &  357.3  & -54.75\\
% 2  &  356.4  & -56.25\\ 
% 3  &  353.55 & -52.6\\ 
% 4  &  353.55 & -54.35\\ 
% 5  &  353.3  & -56.25\\ 
% 6  &  350.4  & -52.6\\ 
% 7  &  349.8  & -54.75\\ 
% 8  &  350.1  & -56.25\\
% 9  &  356.1  & -53.5\\ 
% 10 &  355.2  & -55.0\\ 
% 11 &  352.0  & -53.5\\
% 12 &  352.0  & -55.0\\
% 13 &  348.8  & -53.5\\
%\hline\hline
%\end{tabular}
%\end{center}
%\label{tab:fieldCentres}
%\end{table}

\subsection{Data processing}

After each run, we processed the raw data with version 5.35 of the
{\tt 2dfdr}
pipeline\footnote{http://www.aao.gov.au/science/instruments/AAOmega/reduction}. Redshifts
were then measured (see Sec.~\ref{sec:redshifts}), and all objects
with secure redshifts were removed from the target catalogue before
the next run. This maximised the observing efficiency in the second
and third observing runs. After the first run, we also modified the pointings shown in
Fig.~\ref{fig:objects} to improve the targeting efficiency.

We have since reprocessed the data with a modified version 6.2 of the
{\tt 2dfdr}
pipeline.  Version 6.2 improves on version 5.35 in several notable
ways. Tram line mapping is more precise, and one can now
simultaneously fit a model of the background scattered light together
with the flux in the fibres to each column of the detector. The fibre
traces are almost parallel to the detector rows.

The further modifications we made to v6.2 of {\tt 2dfdr} improve on the publicly available
version is several key aspects. These improvements include:

\begin{itemize}

\item Using PyCosmic \cite{Husemann2012} to find pixels affected by cosmic
  rays. These pixels are removed from the analysis.

\item Using singular value decomposition (SVD) when the least-squares
  solution fails to produce an adequate solution to the matrix
  equation that is used to optimally extract the flux in the fibres
  (see Sharp \& Birchall \shortcite{Sharp2010} for details). On
  average, we use SVD instead of least squares once every 200
  columns.

\item For low signal-to-noise regions in the fibre flats (typically in
  the very blue end of the spectrum), average over 10 columns when
  fitting the fibre profile.

\item Using the {\tt 2dfdr} provided 2d background subtraction routine instead of the
  1d sky subtraction routine to remove the scattered light background
  before fitting the fibre profiles.

\end{itemize}

These modifications will be incorporated in future versions of {\tt 2dfdr}.

Once processed, we then splice the red and blue ends of the spectra. Since some
targets are observed over different nights and can appear in more than
one field (because of field overlap) we sum all the data on a single
target into one spectrum, weighting on the signal-to-noise ratio.

These improvements lead to better quality data. As a result, we get
about 10\% more redshifts than we did with version 5.35 of {\tt
  2dfdr}.  While the data are better, there is still room for
improvement. In particular, one can sometimes see a discontinuity in
the spectra near to wavelength where the dichroic splits the light
into the two arms
of the spectrograph. This is largely due to errors in sky subtraction
and is caused by a combination of residual background scattered light,
errors in estimating the normalisation of the fibre throughputs, and
poor spectral and spatial uniformity in the fibre flats. The accuracy
of the sky subtraction, which is assessed by examining how well the
sky is removed from the sky fibres, is about 1\% in the continuum.
%We are actively 
%investigating ways to improve the accuracy.

\section{Redshift Estimates\label{sec:redshifts}}

We used {\tt runz} \cite{Drinkwater2010} to inspect each spectrum, to
measure a redshift, and to note down any relevant aspects, such as the
presence of broad emission lines. For each spectrum, we assign a
quality flag that varies from 1 to 6. The flags are identical to those
used in the OzDES redshift survey \cite{Fang2015}, and have the
following meanings.

\begin{itemize}

\item 6 - a star
\item 4 - $> 99$\% redshift consistency in overlap regions
\item 3 - $\sim 95$\% redshift consistency in overlap regions
\item 2 - an educated guess
\item 1 - unknown

\end{itemize}

Only objects with flag 3, 4 and 6 are listed in our catalogue. Flag 5 is an
internal flag used by OzDES, and is not used here. The fraction of
objects with the flag set to 6, 4, or 3, or less than 3 were 0.10, 0.53, 0.12 and
0.25, respectively.

\begin{figure*}
\begin{center}
\includegraphics[width=40pc]{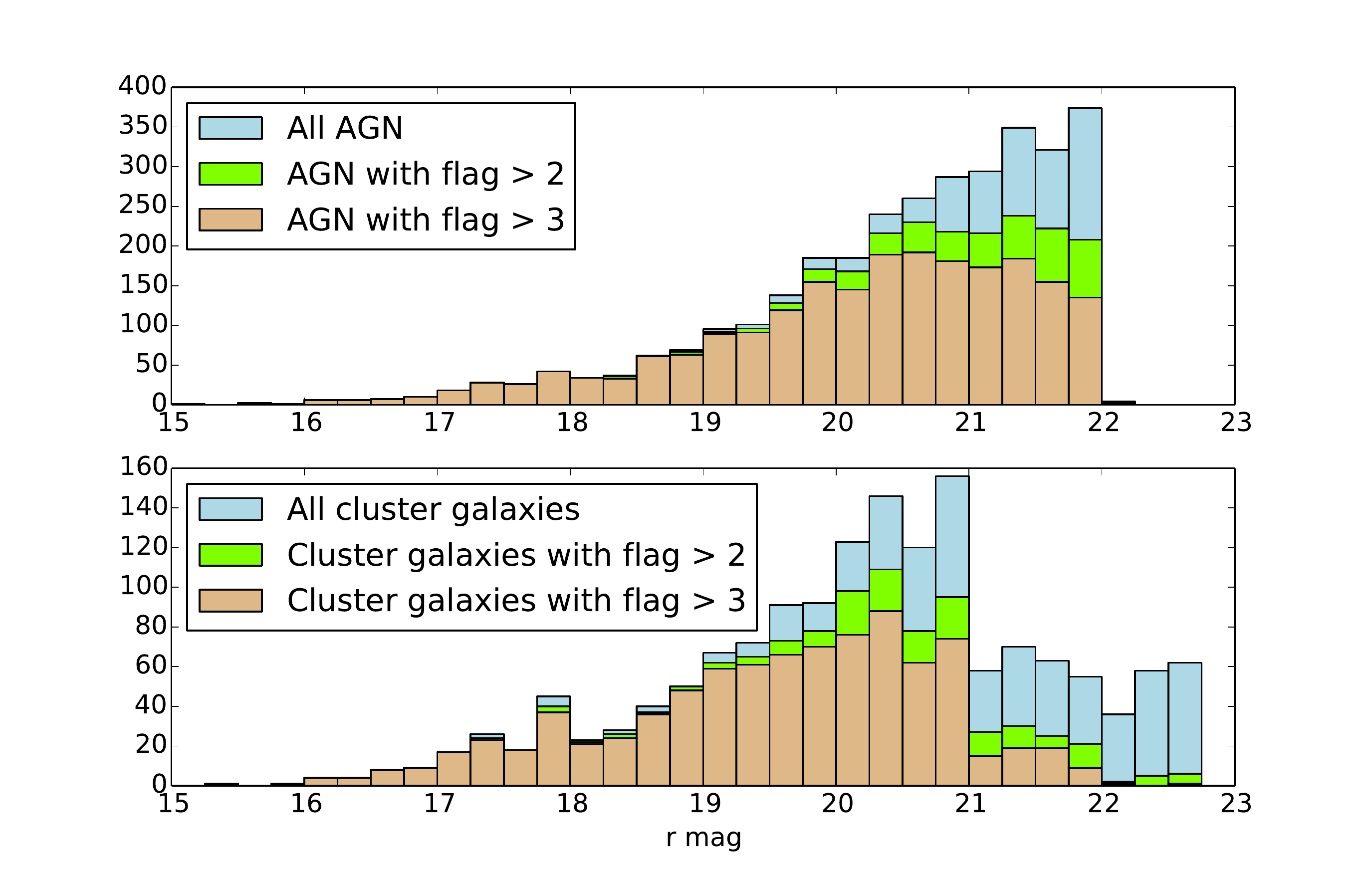}
\caption{Redshift completeness as a function of r band magnitude for AGN
  (above) and cluster galaxies (below).}
\label{fig:completeness}
\end{center}
\end{figure*}

All redshifts are placed in the heliocentric reference frame. The
correction is small, usually less than 0.0001 in redshift. The
redshift completeness as a function of magnitude and the redshift
quality flag is shown in Fig.~\ref{fig:completeness} for AGN and
cluster galaxies.

\subsection{Redshift uncertainties}

The considerable overlap between fields (see Fig.~\ref{fig:objects})
means that about 700 objects were observed more than once. We noted
earlier that we had processed the data twice, once with an earlier
version of 2dfdr and again with a revised version. On the first
occasion, we measured redshifts of objects in each field independently. By
comparing the redshifts of objects in the overlap regions, we get an estimate of redshift uncertainties and
an estimate of the number of times the redshifts agree.
% Need to estimate how many objects were observed more than once.
%We have also measured redshifts for the entire dataset twice. Once in 2013
%when the data were taken, and once more in 2015 when the data were
%reprocessed with the latest version of {\tt 2dfdr} pipeline.

For objects with the flag set to 4 or 6, there were
  no inconsistent redshifts.  For objects that have the
flag set to 3, the percentage of objects with consistent redshifts was 95\%.

We use these repeat observations to also measure the RMS difference
between pairs of observations. We estimate the redshift uncertainty by dividing
the RMS by $\sqrt{2}$.  The uncertainty depends on object type. For
AGN, which generally have broad lines, the uncertainty is typically
$0.001 (1+z)$. For galaxies, the corresponding uncertainty is $0.0002
(1+z)$. %The redshifts of stars are
%set to zero.

\subsection{The redshift catalogue and object demographics}

The first four lines of the catalogue are shown in
Table~\ref{tab:redshiftCatalogue}. The full catalogue can be obtained
from CESAM\footnote{http://cesam.oamp.fr/xmm-lss/}, via the XXL Master
Catalogue Browser\footnote{http://cosmosdb.iasf-milano.inaf.it/XXL}, or by contacting
the first author of this paper.

\begin{table*}
\caption{The first four objects in the redshift catalogue}
\begin{center}
\begin{tabular}{llllcc}
\hline\hline
Name & R.A. (J2000) & Dec. (J2000)& Redshift & Redshift flag & Comment \\
& (deg) & (deg) &  &  & \\
\hline
XXL-AAOmega J234817.22-541456.6  &  357.07175   &    -54.24905 &   2.2079 &  4  & AGN \\
XXL-AAOmega J234853.07-534134.0  &  357.22113   &    -53.69278 &   0.6792 &  4  & AGN \\
XXL-AAOmega J234926.54-534618.2  &  357.36060   &    -53.77172 &   1.6434 &  4  & AGN \\
XXL-AAOmega J234937.79-534724.9  &  357.40747   &    -53.79025 &   0.2560 &  4  & AGN \\
\hline\hline
\end{tabular}
\end{center}
\label{tab:redshiftCatalogue}
\end{table*}

%\begin{sidewaysfigure*}
%\begin{landscape}
\begin{figure*}
\begin{center}
\includegraphics[width=40pc]{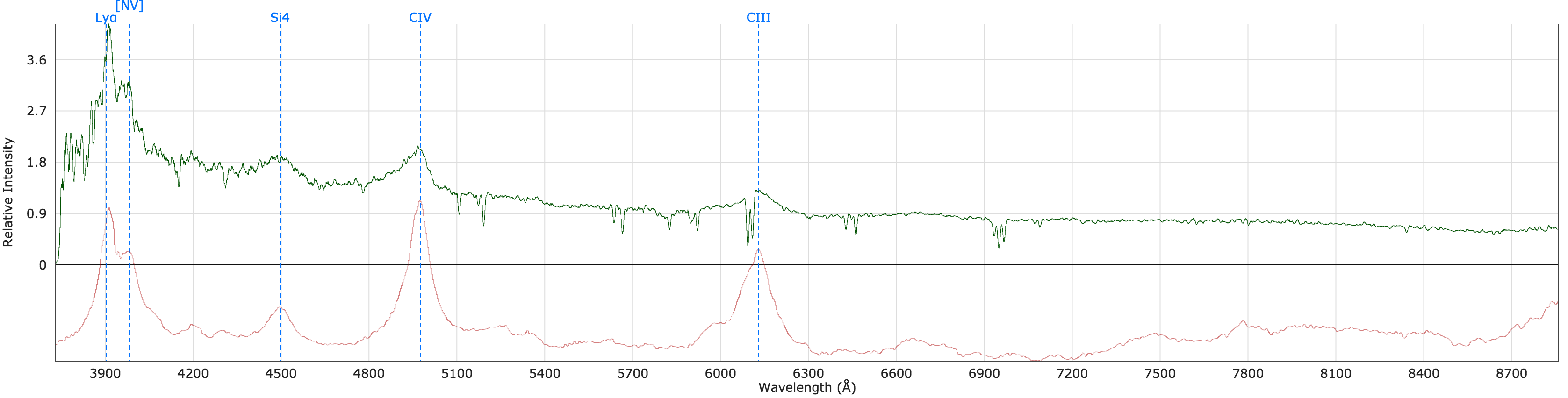}
\includegraphics[width=40pc]{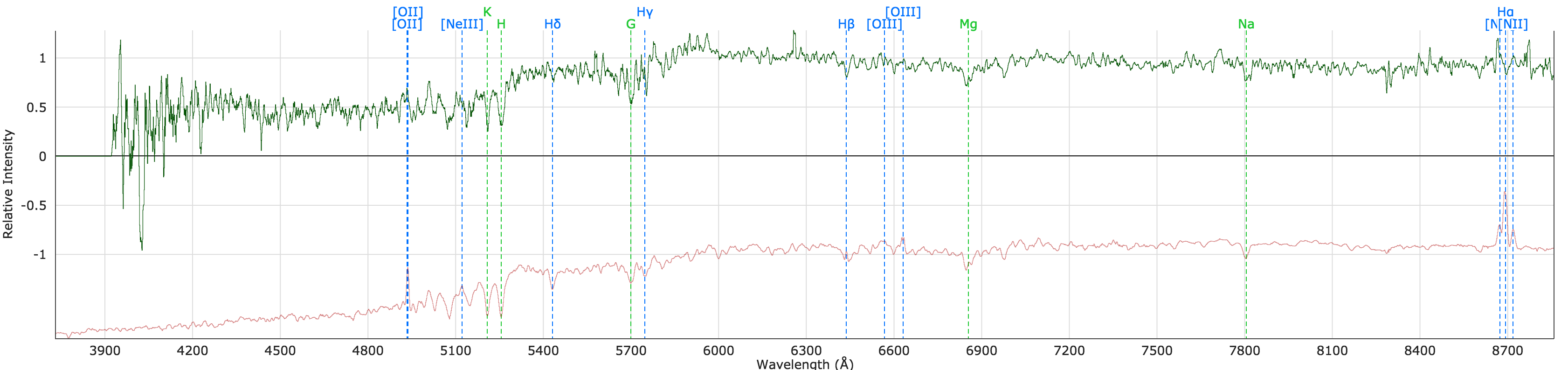}
\caption{Sample spectra (in green) showing an AGN with broad emission lines
  (above) and a cluster
galaxy (below). Template spectra are shown in red. The vertical lines
mark commonly observed lines in AGN
and galaxies. This plot was
produced using the output from {\tt Marz} (Hinton et al. in preparation).}
\label{fig:spectra}
\end{center}
\end{figure*}
%\end{landscape}
%\end{sidewaysfigure*}

The comment field is usually blank or has one of two values: AGN or
binary.  If broad AGN-like features are visible in the spectrum, then
we mark the object as an AGN (see Fig.~\ref{fig:spectra} for an
example of an object with broad emission lines). About half of the 3,184 targeted X-ray
point sources have broad AGN-like features. Redshifts for about half
of the sources that remain (i.e. those without broad lines) could be
obtained from galaxy lines. The remaining sources, about 25\% of the
total, do not have a redshift. About 7\% of the X-ray point sources
were stars (the redshift quality flag is set to 6 for such objects). A
detailed breakdown of the number of redshifts obtained is provided in
Table~\ref{tab:redshiftSummary}.

The comment 'binary' means that the spectral features of two stars,
usually an M dwarf with a white dwarf, are visible in the
spectrum. There are three of these in the catalogue.

Of the 1,674 objects targeted as candidate cluster galaxies, 830 have
redshifts. Broad AGN-like features could be seen in 26 of the spectra.
Another 278 objects are stars, of which nearly all are M dwarfs.

\begin{figure}
\begin{center}
\includegraphics[width=20pc]{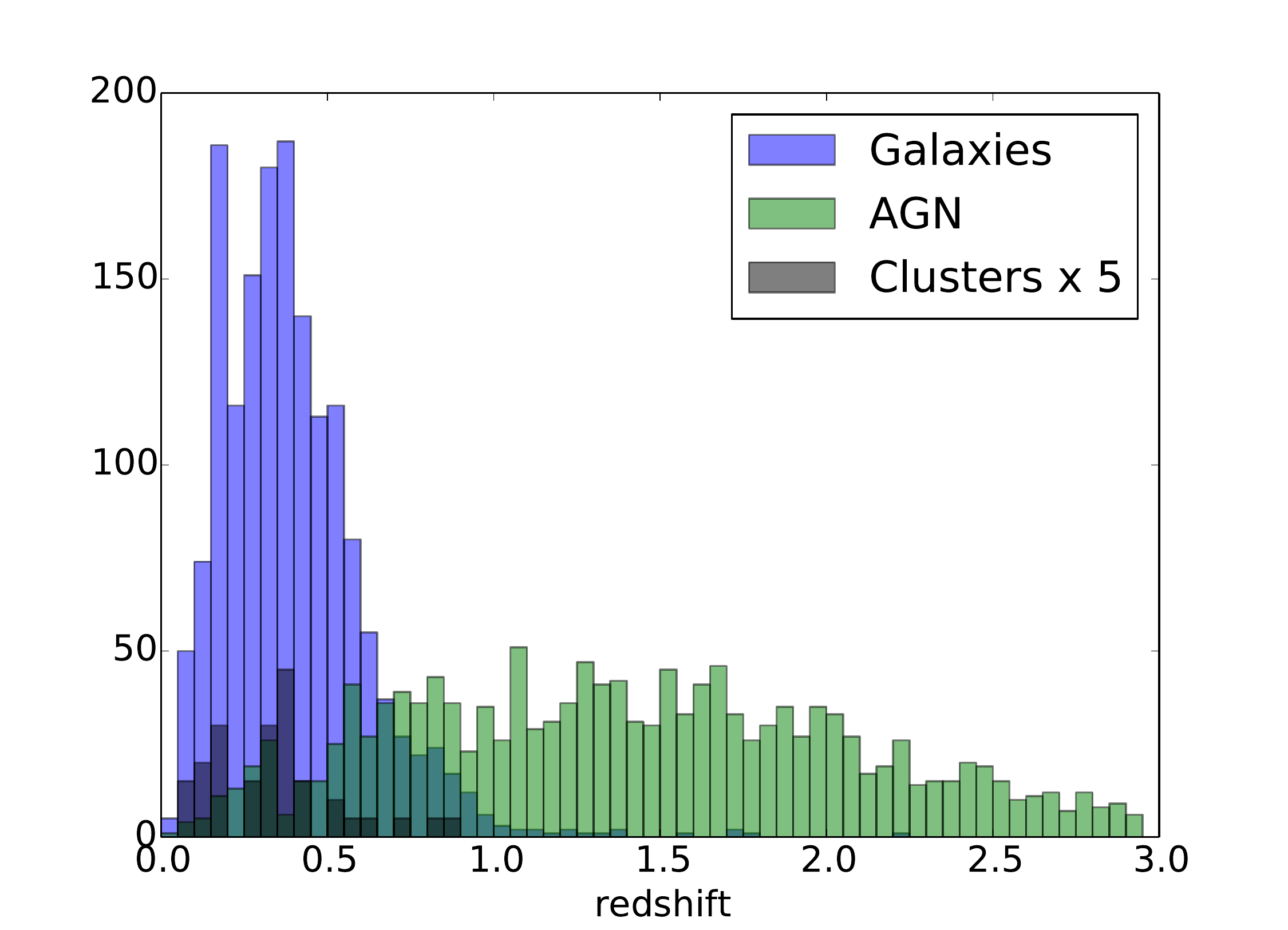}
\caption{A redshift histogram of AGN, clusters (scaled by x5) and galaxies in
  the XXL southern field. The AGN extend out to $z\sim5$, and the
  clusters out to $z\sim0.9$. Not shown are the 44 AGN in the redshift
interval $3 \le z \le 5$.}
 \label{fig:redshiftDistribution}
\end{center}
\end{figure}

In Fig.~\ref{fig:objects}, we show the distribution of objects with
redshifts on the sky, and in Fig.~\ref{fig:redshiftDistribution}, we
show the redshift distribution of AGN, and galaxies. Galaxies are
largely restricted to be below $z\sim1$, whereas AGN go out to
$z\sim5$. 

Also shown in Fig.~\ref{fig:redshiftDistribution} is the redshift
distribution (scaled by x5) of 41 galaxy clusters whose redshifts
were measured through these observations. These 41 clusters represent
a subsample of the 100 most X-ray luminous clusters (49 of which are
in XXL-South\footnote{The redshifts of the remaining 8 clusters have
  been or are currently being obtained with other facilities. See
  XXL Paper II for details.}) listed in XXL Paper II. For this subsample, we targeted
838 objects within 1\,Mpc of the cluster X-ray centres. Of these, we
obtained redshifts for 501 galaxies, of which 211 were found to belong
to the cluster.

\begin{table*}
\caption{The number of objects targeted and the number of redshifts obtained}
\begin{center}
\begin{tabular}{lrrrr}
\hline\hline
Object & Objects Targeted & Objects with redshifts & Stars & Objects with broad AGN-like features \\
&& (excludes stars) &&\\
\hline
AGN                        & 3,184 & 2,332  & 220 & 1,533 \\
Cluster Galaxies     & 1,674  &    830  & 278 & 26 \\
\hline
Total                      & 4,858 & 3,162 & 498 & 1,559 \\
\hline\hline
\end{tabular}
\end{center}
\label{tab:redshiftSummary}
\end{table*}
 
% Figure of the redshift distribution ...

\section{Velocity Dispersions}

The nearest clusters have the largest angular extent on the sky, so it
was sometimes possible to allocate more than one fibre to cluster
members in a single plate configuration. Sometimes this was as high as
five. We were also able to free up fibres in later runs once 
a redshift was obtained, and there is quite some overlap between the 13 AAOmega
pointings. Altogether, this meant that some of the nearer clusters have over a
dozen spectroscopically confirmed members from the AAOmega
spectroscopy alone.  This is sufficient to measure a velocity
dispersion.

Using the location of the 49 brightest X-ray selected clusters listed
in XXL Paper II, we select galaxies as potential cluster members if they
satisfy the following criteria. They have more than 10 members with
redshifts that land within $0.01*(1+z_{cl})$ of the cluster redshift
$z_{cl}$ and have positions that land within 1\,Mpc of the cluster
centre. There are 8 such clusters in total. From these members, we use
either the biweight midvariance or the gapper \cite{Beer1990} as a
robust estimator of the velocity dispersion. The former is used if
there are more than 15 objects; while the latter is used
otherwise. For our sample, there is little difference between the two
methods.  The results are listed in Table~\ref{tab:velDisp}. The
uncertainties of both methods are estimated using bootstrap
resampling. The uncertainties in the redshifts
  translate to an uncertainty of $\sim 80$\,km/s at $z\sim0.3$. The
  impact of these uncertainties on the velocity dispersions
  that are computed for these clusters is negligible.

The X-ray centroids of XLSSC 535 and XLSSC 536, which are the X-ray
counterparts of Abell\,4005, are separated by 3.5
arc minutes, which corresponds to 600\,kpc at z=0.17.  The 24
spectroscopic members observed do not show evidence of a bimodal
distribution in redshift. Hence, we consider them together, when
measuring the velocity dispersion.

There is a positive correlation between the velocity dispersions
listed in Table~\ref{tab:velDisp} and
the X-ray determined masses that are listed in XXL Paper II; however,
given the size of the sample (only 8 clusters) and the large uncertainties in both the
mass and the velocity dispersion, the correlation is not significant.

\begin{table}
\caption{Cluster velocity dispersions for 8 of the 49 clusters in
  XXL Paper II.}
\begin{center}
\begin{tabular}{lccl}
\hline\hline
Cluster & z & Members & Vel. disp. \\
&&&[km/s] \\
\hline
XLSSC 535/536$^a$    & 0.171 & 24 & $660^{+90}_{-110}$\\
XLSSC 524$^{b}$          & 0.270 & 11 & $640^{+100}_{-200}$\\
XLSSC 538                   & 0.332 & 13 & $1080^{+170}_{-320}$\\
XLSC 544                   & 0.095 & 11 & $530^{+100}_{-200}$\\
XLSSC 545                   & 0.353 & 12 & $370^{+20}_{-80}$\\
XLSSC 547                   & 0.367 & 16 & $810^{+100}_{-160}$\\
XLSSC 548                   & 0.321 & 14 & $800^{+60}_{-180}$\\

\hline\hline
\end{tabular}
\end{center}
\label{tab:velDisp}
\tabnote{$^a$ The X-ray counterpart to Abell\,4005}
\tabnote{$^b$ Cluster 511 in \u{S}uhada et al. \shortcite{Suhada2012} }
%\tabnote{$^c$ Cluster  in \cite{Menanteau2010}} 

\end{table}

\section{Summary}

Using AAOmega on the AAT, we have obtained the redshifts of 3,660
X-ray selected sources in the XXL southern field.  Of these 3,660
sources, 1,558 are broad line AGN, and 498 are stars. The remaining
objects are either galaxies from the AGN sample without broad lines,
field galaxies, or cluster galaxies in XXL detected clusters.

Of the 49 brightest X-ray selected clusters in the XXL southern field,
we get redshifts for 41 of them.  We compute the velocity dispersions
for clusters with more than 10 members. The velocity dispersions are
indicative of clusters that range from rich groups to massive
clusters.

%The full catalogue can be obtained
%from CESAM\footnote{http://cesam.oamp.fr/xmm-lss/}, via the XXL Master
%Catalogue Browser\footnote{http://cosmosdb.iasf-milano.inaf.it/XXL}, or by contacting
%the first author of this paper.

% UNCOMMENT THE LINES BELOW IF YOU WISH TO USE BIBTEX
%\bibliographystyle{apj}
%\bibliography{yourbibfile}

\begin{acknowledgements}
  Based in part on data acquired through the Australian Astronomical
  Observatory, under programs A/2013A/018 and
  A/2013B/001. O.M. is grateful for the financial
    support provided by the NEWFELPRO fellowship project in
    Croatia. F.P. acknowledges support from the BMBF/DLR grant 50 OR
    1117, the DFG grant RE 1462-6 and the DFG Transregio Programme
    TR33.
\end{acknowledgements}

%\bibliography{mybib}{}

\end{document}